\documentclass[aps,twocolumn,amsfonts,amsmath,amssymb,nofootinbib,longbibliography,floatfix]{revtex4-2}

\setcitestyle{round,authoryear}

\usepackage{mathtools}
\usepackage{graphicx}

\usepackage{xcolor}

\usepackage{url}

\usepackage[breaklinks]{hyperref}
\usepackage[hyphenbreaks]{breakurl}

\usepackage{enumerate}
\usepackage{enumitem}

\usepackage{psfrag}

\newtheorem{theorem}{Theorem}

\newtheorem{definition}{Definition}
\newtheorem{remark}{Remark}

\begin{document}


\title{A mathematical construction of the \( 260 \) day
mesoamerican calendar based on archaeoastronomical alignments
}

\author{Sergio Mendoza}
\email[Email address: ]{sergio@astro.unam.mx}
\affiliation{$^1$Universidad Nacional Aut\'onoma de M\'exico, Instituto
de Astronom\'{\i}a, AP 70-264, Ciudad de M\'exico 04510, M\'exico
             }


\keywords{Mesoamerica -- Calendar -- Gravitation -- Mathematics: Arithmetic --
Arqueoastronomy }

\begin{abstract}
  Ancient mesoamerican cultures built a short ritual \( 260 \) day
calendar and used it for daily routinary life.  Using simple arithmetic
calculations it is first shown that by forcing the introduction of the the
fundamental number \( 13 \) to calculate days in a calendar, a \( 364 \)
day count can be built and from this, the short mesoamerican calendar
of \( 260 \) days is constructed.  It is also shown that the basic
mesoamerican relation between the full solar \( 365 \) day calendar and
the short one of \( 260 \) days given by: \( 365 \times 52 = 260 \times 73
\) is Kepler's third law of orbital motion between Earth's period of time
about the Sun and an imaginary synodic orbit with a \( 260 \) day period.
Based on this basic mesoamerican relation and a general definition of an
archaeoastronomical alignment, a full mathematical definition of a short
calendar count is made.  For particular cases, approximate calendar counts
of \( 364 \) and \( 360 \) days are obtained with fundamental numbers \(
13 \) and \( 18 \) associated to each one respectively.  The extra days
required for this approximate calendar counts are a given solstice day
for the former, and \( 4 \) plus \( 1 \) (the given solstice day) for the
latter, generating the \( 5 \) nemontemi days used by ancient mesoamerican
cultures.  Motivated by an extended mesoamerican relation, two types of
archaeoastronomical alignments are defined and their consequences are
investigated.  These Type-A and Type-B archaeoastronomical alignments
motivate the introduction of extended fundamental or monad fractions
that partition the approximate and short counts.
\end{abstract}

\maketitle
\newpage

\section{Introduction}
\label{introduction}

  Mesoamerican cultures used a \( 260 \) day short
or ritual calendar, combined with a \( 365 \) day solar year
calendar~\citep[e.g.][]{native-math,ortiz-franco,thompson2017,harvey-williams80,introduction-cultural-maths}.
The use of a base \( 20 \) numerical system for counting, shines a light
on the relevance of the number \( 260 \), since \( 260 = 20 \times 13
\) and the number \( 13 \) was considered very important, of religious
significance~\citep[see e.g.][]{aztec-thought,diosesprehispanicos}.
Thus, in principle it seems that \( 260 \) represents arithmetical
harmony and the discussion about this choice should end here.  However,
one can argue as to why not use a shorter calendar of say \( 130 = 260 /
2 \) days or a larger one of \( 520 = 260 \times 2 \) days, etc.

  The \( 260 \) days of the short mesoamerican calendar  can be
accounted for using the fact that the band located between \( 14^\circ
42' \text{N} \) and \( 15^\circ \text{N} \) where the Sun crossing the
Zenith occurs close to August 12--13  and \( 260 \) days later once
more in April 30 -- May 1.  At this latitude there is a \( 260 \) day
time interval between the northern autumn and spring zenithal transits
of the sun~\citep[cf.][and references therein]{cosmology-calendar}.
\citet{henderson74} had the concern that there is a complementary \(
105 \) day interval between both dates which is not divisible by the
relevant mesoamerican number \( 52 \), unless an added calibration of
one day every solar year cycle was performed due to the fact that \(
105 \) is \( 104 = 52 \times 2 \) plus \( 1 \).

  This article presents a mathematical approach for the construction
of the \( 260 \) day mesoamerican short calendar system based on
archaeoastronomical alignments about a given solstice.  The study begins
by exploring and trying to fit the fundamental mesoamerican number \(
13 \) into a full solar year of \( 365 \) days.  In a natural way,
this leads to a construction of an ``approximate'' solar year of
\( 364 \) days and eventually to the \( 260 \) day short calendar
creation.  In Section~\ref{numbers} the whole arithmetic study is
produced using the prime number decomposition of the numbers \( 365
\), \( 260 \) and \( 364 \), based on some numerical similarities
and symmetries.  In Section~\ref{kepler} it is noted the fact that
Kepler's third law of planetary motion in synodic coordinates yields
the basic mesoamerican relation \( 260 \times 73 = 365 \times 52 \).
In Section~\ref{archaeoastronomy} it is  shown the relevance of
the \( 365 \), \( 364 \) and \( 260 \) mesoamerican counts made in
two archaeoastronomical sites  and it is noted that basic fractions
or monads were used to divide the \( 365 \) and \( 364 \) counts.
Section~\ref{venus} shows that using the synodic period of Venus and
forcing the number \( 13 \) to enter into Kepler's third law, the short
\( 260 \) day calendar is also obtained.  In Section~\ref{uniqueness}
it is developed a formal mathematical way of building an approximate
solar calendar with a natural creation of a short calendar count,
based on two symmetrical archaeoastronomical alignments occurring about
a given solstice.  This definition has a lot of consequences and it is
able to account for different calendar counts, including the \( 360 +
5 \) days count used by ancient mesoamericans.  It also shows how the
number \( 13 \) is a consequence of an approximate solar calendar of \(
364 \) days, but its relevance is not unique since for this approximate
solar calendar, number \( 18 \) becomes as fundamental as \( 13 \) related
to the approximate and short counts.  The whole analysis also suggest
extended fundamental monad fractions Finally in Section~\ref{discussion}
a discussion of the obtained results is made.

\section{The importance of number 13}
\label{numbers}

  The prime number decomposition of \( 365 \) and \( 260 \) are
respectively given by:

\begin{gather}
  365 = 73 \times 5,
  	\label{prime365}  \\
  260 = 52 \times 5 = 2^2 \times 13 \times 5.
  	\label{prime260}
\end{gather}

\noindent It is immediately evident then that the the first choice for
the number \( 260 \) is that it is divisible by 5, a characteristic
inherited by the \( 365 \) days in a solar year.

  As mentioned in the introduction, the number \( 13 \) had a
strong relevance in mesoamerican cultures and as described before  it
appears in the prime number decomposition of \( 260 \) as expressed in
equation~\eqref{prime260}.

  The prime decomposition of \( 364 \), which is the number of
days of a solar calendar minus one, is:

\begin{equation}
  364 = 7 \times 52 = 2^2 \times 7 \times 13.
\label{prime364}
\end{equation}

\noindent From this relation it can be seen that the number \( 364 \) is
divisible by \( 2 \), \( 2^2 \), \( 13 \) and \( 52 \),  a property shared
by the number \( 260 \).  Also, the number
\( 52 \) divides \( 260 \) into \( 5 \) parts and \( 364 \) into \( 7 \),
so that:

\begin{equation}
  \frac{ 364 }{ 7 } = \frac{ 260 }{ 5 } = 52, \quad \text{i.e.:} \quad 
    364 \times 5 = 260 \times 7 = 1820.
\label{tempo-relation}
\end{equation}

\noindent As it will be discussed in Section~\ref{archaeoastronomy}, some
mesoamerican cultures aligned their monuments using this property.

  The prime number decomposition of integers \( \lesssim 365 \) \citep[see
e.g.][]{wikipediaprimes} that contains the fundamental number \( 13 \)
is exactly \( 364 \).  So, if the number \( 13 \) is to appear in the
prime number decomposition of an approximate number of days for the
solar calendar, then \( 364 \) is to be chosen, with the added bonus
that the rectification to get the correct number of \( 365 \) days in a
solar year is by waiting or adding one extra day after the \( 364 \)th
day each cycle\footnote{In a somewhat related manner, the accumulated
extra hours produced by the \( 365 \) day count in a full solar year
(accounted for leap years in the Gregorian calendar)  could have been
corrected by mesoamerican cultures, but this issue is
controversial and has been strongly debated over the years~\citep[cf.][and
references therein]{ezcurra22,prajc2023,ezcurra2023}.}

  Since \( 364 \) is obtained after \( 7 \) periods of \( 52 \) according
to equation~\eqref{prime365}, if a short calendar is to be constructed it
seems reasonable to use a shorter (\( < 7 \)) number of periods.  So, it
seems that a good choice for the number of days in a short calendar could
be any of the following:

\begin{align*}
  364 &= 52 \times 7 = 13 \times 2^2 \times 7  = 13 \times 2^2 \times 7,
  	\\                                     
  312 &= 52 \times 6 = 13 \times 2^2 \times 6  = 13 \times 2^3 \times 3,
  	\\                                     
  260 &= 52 \times 5 = 13 \times 2^2 \times 5,
  	\\                                     
  208 &= 52 \times 4 = 13 \times 2^2 \times 4 = 13 \times 2^4,
  	\\                                     
  156 &= 52 \times 3 = 13 \times 2^2 \times 3,  = 13 \times 2^2 \times 3,
  	\\                                     
  104 &= 52 \times 2 = 13 \times 2^2 \times 2,  = 13 \times 2^3,
  	\\                                     
   52 &= 52 \times 1 = 13 \times 2^2.
\end{align*}

  From the previous set of numbers it can be seen that all choices
could represent \( 13 \) days, grouped into sets of:
\( 28,\ 24,\ 20,\ 16,\ 12,\ 6,\ 4\) respectively.   One can argue here
that the number \( 28 \) is quite close to the sidereal period of the
moon but there is no evidence that mesoamerican cultures knew about the
sidereal concept of orbiting objects in the celestial sphere.  In any case,
\( 28 \) is quite close to the number of days of the synodic period of
the moon which is \( 29.5 \) days.   A closer look at the prime
decomposition shown in the previous set of equations shows that the number
\( 260 \) is the only one in the set which is divisible by \( 5 \), 
a property shared with the number \( 365
\) as shown in equation~\eqref{prime365}.  In other words:

\begin{equation}
  \frac{ 365 }{ 73 } = \frac{ 260 }{ 52 } = 5,
\label{golden-rule}
\end{equation}

\noindent which implies:

\begin{equation}
  365 \times  52 = 260 \times 73 = 18980.
\label{fuego-nuevo}
\end{equation}

\noindent This is the basic mesoamerican calendar relation between the
complete solar cycle and the short one, since if both counts start at
the same time it takes \( 18980 \) days for them to coincide again.
Apart from the remarkable property of equation~\eqref{golden-rule},
the number \( 260 \) is the only one on the list which is divisible
by \( 20 \), the base number used in ancient mesoamerica.  As such, \(
13 \) days grouped in sets of \( 20 \) or ``veintenas'' can be chosen.
As a direct consequence of this, it follows that the number \( 260 \) is
also divisible by \( 4 \), so the \( 260 \) day count can also be grouped
into quarters.  Each of these four periods have \( 65 \) days each since
\( 260 = 65 \times 4 \). Having a division by \( 4 \) on the calendar
was also performed in the full solar year, since formally this solar year
consists of \( 360 \) normal days plus \( 5 \) special ``nemontemi'' days.
In this case, \( 360= 4 \times 90 \) is naturally divisible by \( 4 \)
with a resulting \( 90 \) days on each of the four periods. Since \( 360 /
20 = 18 \), the mesoamericans constructed a solar year calendar consisting
of \( 18 \) sets each having \( 20 \) days, plus \( 5 \) nemontemi days.
The \( 364 \) day count can also be divided into \( 13 \) sets, each
containing \( 28 \) days.  This approximate solar year has exactly \( 4 \)
important divisions of \( 91 \) days each.  Note that this approximate
solar year cannot be divided into periods of \( 20 \) days, nor is
divisible by \( 5 \).  But its relevance and use on mesoamerican cultures
through archaeoastronomy is discussed in Section~\ref{archaeoastronomy}.
It will also be the base in Section~\ref{uniqueness} to make a formal
definition of a short calendar count.

  An  alternative way to choose the \( 260 \) day calendar
from the approximate solar \( 364 \) cycle is obtained in the following
manner.  According to equation~\eqref{prime364}: \( 364 = 13 \times 28 \).
If the requirement is to find a shorter calendar that contains the number \( 13
\) in its prime number decomposition which also includes
the prime number \( 5 \), sharing
this property with the full solar year of \( 365 \) days, then the numbers:

\begin{gather*}
  325 = 13 \times 25, \qquad 260 = 13 \times 20,  \\
  195 = 13 \times 15, \qquad 130 = 13 \times 10,
\end{gather*}

\noindent are good candidates.  However, the only number divisible by \( 20
\) from the previous list is \( 260 \), with the extra property that is
also divisible by \( 4 \).

  The introduction of the number \( 13 \) on the full solar year of \( 365
\) days can be performed using a different approach 
by the multiplication of these numbers and so,
using the prime number decompositions of equations~\eqref{prime365}
and~\eqref{prime364} it follows that:

\begin{equation}
  \frac{ 365 \times 13 }{ 5 \times 73} = \frac{ 364}{ 2^2 \times 7 } = 13,
\label{t00}
\end{equation}

\noindent i.e.:

\begin{gather*}
  365 \times \left( 2^2 \times 13 \right) \times 7 = 364 \times 5 \times
    73,  \\
\intertext{so that:}
  \begin{split}
  \left( 365 \times 52 \right) \times 7 &= 364 \times 5 \times
    73,  \\
			&= \left( 2^2 \times 7 \times 13 \right) \times 5
			\times 73, \\
			&= \left( 20 \times 13 \right) \times 73 \times 7,
			\\
                        &= \left( 260 \times 73 \right) \times 7,
  \end{split}
\end{gather*}

\noindent which after division by \( 7 \) yields the mesoamerican
basic calendar relation~\eqref{fuego-nuevo}.

\section{Kepler's third law}
\label{kepler}

  There is no evidence that the mesoamerican cosmological culture
identified the motion of some of the bright points on the celestial
sphere as planets orbiting about the sun.  This is the reason as to why
all orbital motions of celestial bodies observed by the mesoamerican
cultures are to be described in synodic coordinates, i.e. coordinates
which are relevant to an observer on Earth (and not sidereal coordinates
which are related to an observer at rest with respect to the Sun).
The relation between synodic and sidereal orbiting periods for planetary
orbits smaller than the orbit of the Earth is: 

\begin{equation}
  \frac{1}{T_\text{sid}} = \frac{ 1 }{ T_\oplus } + \frac{ 1 }{ T },
\label{synodic-sidereal}
\end{equation}

\noindent where \( T \) represents the 
synodic period of time, \( T_\oplus = 365  \) days  is the Earth's
sidereal
period of time about the sun and \( T_\text{sid} \) is the 
corresponding sidereal period of time~\citep[see e.g.][]{smart}.

  Since Kepler's third law for circular motion\footnote{The eccentricity of
all planets in the solar system (except for Mercury) 
is quite close to \( 0 \) and so, their orbits are very close to a
circle.} is given by~\citep[e.g.][]{goldstein}:

\begin{equation}
  T_\text{sid} = 2 \pi \sqrt{ \frac{ R^3 }{ G M_\odot } },
\label{kepler-standard}
\end{equation}

\noindent where \( R \) represents the radius of a particular circular
orbit, \( G \) is Newton's gravitational constant and \( M_\odot \) is the
mass of the Sun.  Substitution of equation~\eqref{synodic-sidereal} into
relation~\eqref{kepler-standard} yields:

\begin{equation}
  T = T_\oplus N = ( 365 \, \text{days} ) \times N,
\label{kepler-synodic}
\end{equation}

\noindent where the dimensionless number \( N \) is given by:

\begin{equation}
  N := \frac{ ( R / R_\oplus ) ^{3/2} }{ 1 - ( R/R_\oplus )^{3/2} },
\label{N-tmp}
\end{equation}

\noindent and \( R_\oplus \) is the radius of the orbit of the Earth about
the Sun.  The advantage of Kepler's third law in synodic coordinates as
presented in equation~\eqref{kepler-synodic} is that from the point of
view of an observer on the surface of the Earth, \( N = T / ( \text{365
\,\text{days} )} \), so that if a synodic period of \( 260 \) days is
required --e.g.~the short mesoamerican calendar, then

\begin{equation}
  N = \frac{ 260 }{ 365 } = \frac{ 52 \times  5  }{ 73 \times 5  }  
    = \frac{ 52 }{ 73 }.
\label{N-short-calendar}
\end{equation}

\noindent In the previous equation, the ratio \( 260 / 365 \) has been
expressed as the irreducible fraction \( 52 / 73 \).
Substitution of this result in equation~\eqref{kepler-synodic}
for a value of \( T = 260 \) days yields relation~\eqref{fuego-nuevo}.
In other words, equation~\eqref{fuego-nuevo} is a
representation of Kepler's third law in synodic coordinates, i.e. in
coordinates adapted for an observer of the sky on Earth.  It is important
to note that the dimensionless constant \( N \) was constructed using a
perfect idealised Earth's period orbit of \( 365 \) days and a perfect
idealised orbit of \( 260 \) days.

It is quite extraordinary that the number \( 13 \) was used with so
much power in mesoamerican cultures.  At first sight it all seems to be
related with religion and/or cultural and 
sociological aspects which go beyond the scope of this article\footnote{With respect to cultural aspects,
it is worth mentioning that if the thumb is used as a tool to count the
three phalanxes of each of the remaining four fingers in one's hand,
then number \( 12 \) is reached and to end the counting one still has
the thumb, reaching the prime number \( 13 \).}.  However, we will see in
Section~\ref{uniqueness} that number \( 13 \) was chosen with a fine
tuning mathematical selection for building a short calendar count
motivated by archaeoastronomical alignments.

\section{Some archaeoastronomy examples}
\label{archaeoastronomy}

  There is no better way to show the relevance of the calculations
performed in the previous sections than that of using  two examples of
archaeoastronomy alignments in mesoamerica: Teotihuacan and Tenochtitlan
in Mexico~\citep[see e.g.][for pictures and more Mexican sites with
these properties]{galindo09,galindo10}.

\begin{figure}
  \includegraphics[width=0.47\textwidth]{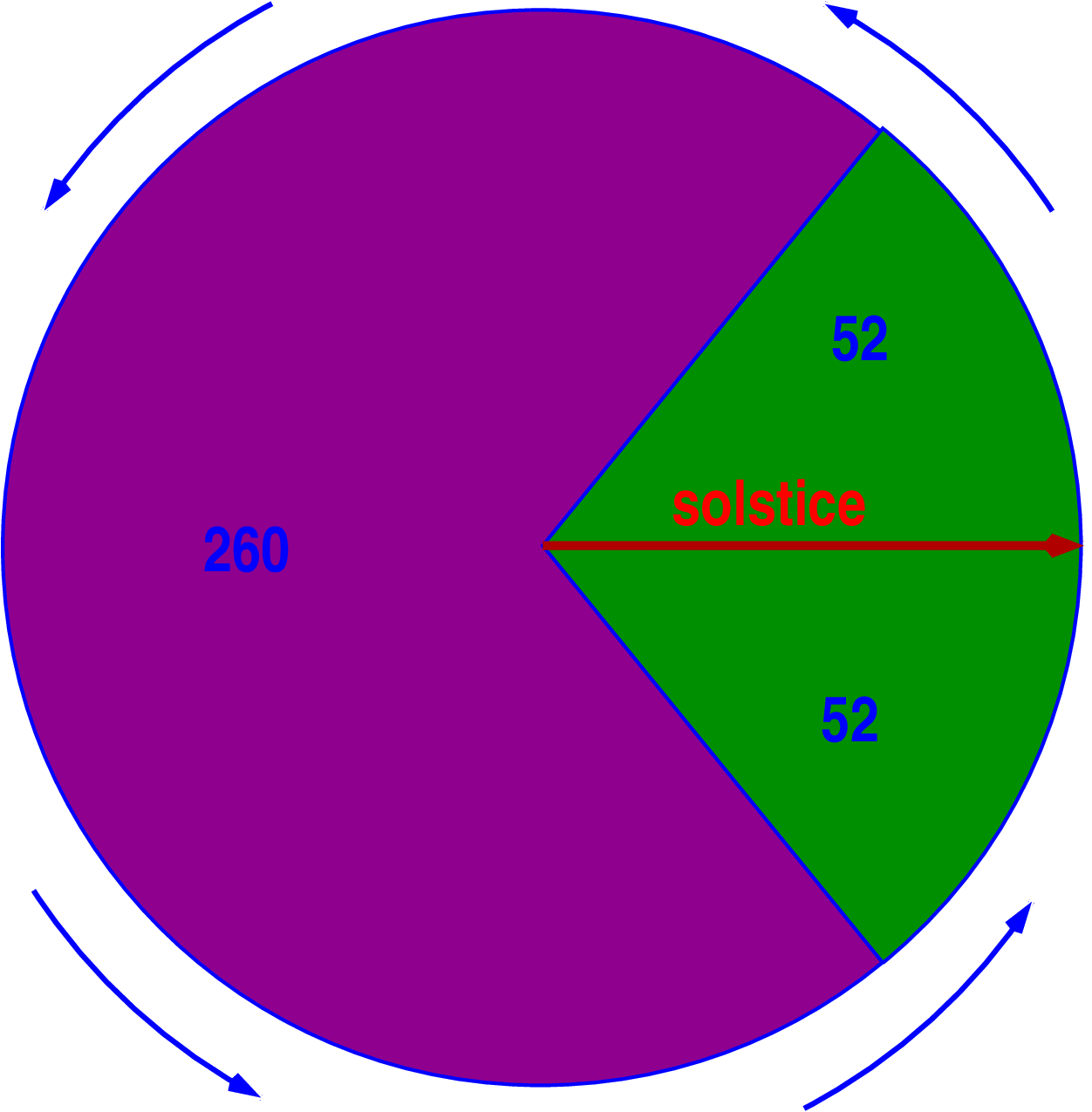}
  \caption{The circular diagram in the Figure represents a \( 365 \)
full solar year count.  The short count of \( 260 \) days represent
the complementary number of days to close after two archaeoastronomical
alignments have occurred \( 52 \) days before and after one given June
or December solstice.  The addition of days is given by \( 260 + 52 +
52 \) plus \( 1 \) extra day which corresponds to the solstice, reaching
the complete full solar year of \( 365 \) days.
}
\label{teotihuacan-figure}
\end{figure}

  The Sun pyramid in Teotihuacan aligns with the solar disc 
on October 29 and on February 12 at sunrise.  The reason for this is that
after the October 29 alignment, one has to wait \( 52 \) days in order to
reach the December 21 solstice.  From this, \( 52 \) days later one arrives
at February 12.  From this last date, after \( 260 = 52 \times 5 
\) days one arrives at the starting point (the next alignment) of 
October 29.  In other words, the first and last day of the short \( 260 \)
count are represented with archaeoastronomical alignments. 
The \( 364 \) day count (which has \( 7 \) periods
of \( 52 \) days each) was used adding an extra day for the solstice.
At sunset, the solar disc aligns with Teotihuacan
Sun's pyramid on April 29 and August 13. From the alignment on April 29 one
has to wait \( 52 \) days to reach the June 21 solstice.  Later on, after 
\( 52 \) days, August 13  is reached.  From this date one has to wait \(
260 \) days to complete the solar year cycle and arrive once more at the
alignment on April 29.
The elapsed time between any pair of these dates divide the approximate
solar year in \( 2 / 7 \) parts and since \( 260 \) divides the same
period of time in \( 5 / 7 \). 
This type of
archaeoastronomical alignment is shown graphically in
Figure~\ref{teotihuacan-figure}.  Note that the extra needed day to
complete the full \( 365 \) day solar year is represented by the given solstice.
In other words, the approximate \( 364 \) day count describes the full
solar year of \( 365 \) days if the given solstice day is added to the 
\( 364 \) count.

\begin{figure}
  \includegraphics[width=0.47\textwidth]{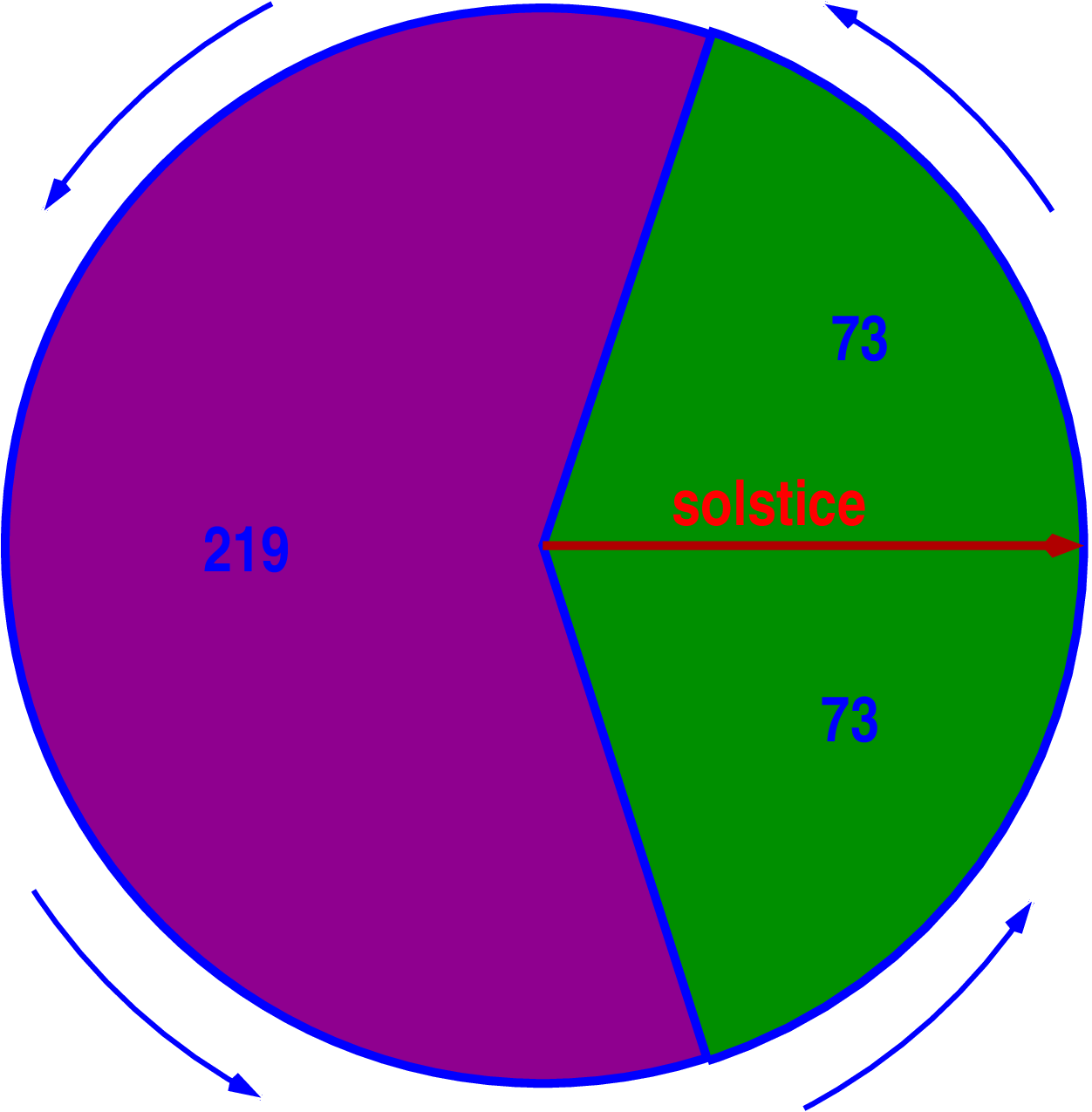}
  \caption{The Figure represents the \( 365 \) days of a full solar year.
Beginning \( 73 \) days before a particular solstice, an archaeoastronomical
alignment is chosen  to occur and \( 73 \) days before the solstice the
alignment repeats once more.  The complete \( 365 \) cycle is finished
after the elapse of \( 219 \) days more.  Note that in this case, as
opposed to Figure~\ref{teotihuacan-figure}, the solstice day needs to be
included on any of the \( 73 \) day intervals.  To resolve the confusion on
this, a coherent understanding of these type of alignments is presented in
Section~\ref{uniqueness}.
}
\label{tenochtitlan-figure}
\end{figure}

  Another method for  archaeoastronomical alignment, using as base not \(
52 \) but \( 73\) was used by the Aztecs in Tenochtitlan, now Mexico City.
The great pyramid of Tenochtitlan (Templo Mayor) aligns with the solar
disc at sunset on April 9 and September 2.  After April 9, one has to
wait \( 73 \) days to reach the June solstice.  From this date, \( 73 \)
days later one reaches September 2.  One has to wait \( 219 = 73 \times
3 \) days from this date to arrive at April 9 again.  At sunrise the
great pyramid of Tenochtitlan aligns with the solar disc on October 9 and
March 4, since after \( 73 \) days from October 9 the December solstice
is reached and from this, \( 73 \) days later one arrives at March 4.
From here, \( 219 = 73 \times 3 \) days later the cycle is completed
arriving at October 9.  The elapsed period of time between any pair of
sunrise or sunset alignment dates divide the full solar year 
of \( 365 \) days in \(
2 / 5 \).  As such, the remaining \( 219 \) days divide the solar year
in \( 3 / 5 \). This type of archaeoastronomical alignment is represented
pictorically in Figure~\ref{tenochtitlan-figure}.
 The problem with this type of arrangement is that the solstice day needs
to be counted inside any of the \( 73 \) day intervals for the alignment
to take place.  We will discuss in detail this in Section~\ref{uniqueness}
and show how the use of the number \( 73 \) is incorrect (but quite close)
for the understanding of these types of archaeoastronomical alignments.

  It is important to remark that  the Aztecs used symbols for some
particular fractions as reported by~\citet{jorgeyjorge}. These symbols
or \emph{monads} correspond to:

\begin{equation}
1/2, \qquad 2/5, \qquad \text{and} \qquad 3/5,
\label{monads-tenochtitlan}
\end{equation}

\noindent of a unit and were represented by an arrow, a heart and a hand
respectively.  The monad fractions \( 2/5 \) and \( 3/5 \) seem to be used
not only for dividing the unit of area, but also for aligning the great
pyramid of Tenochtitlan at sunset or sunrise on some particular dates as
previously discussed.  In any case, both fractions divide the \( 365 \)
days year in a astonishing way and were most probably motivated by the
movement of the solar disc through the full solar year.  From these last
remarks and the ones stated about the division of the archaeoastronomical
alignments at Teotihuacan, it is coherent to postulate that the 
Teotihuacan civilisation used the following 
additional monad fractions:

\begin{equation}
   2/7, \qquad 5/7.
\label{monads-teotihuacan}
\end{equation}

  It will be shown in Section~\ref{uniqueness} that all these monad
fractions and a more general mathematical definition for them can be
constructed once a short calendar count is defined.

\section{Venus}
\label{venus}

  The motion of Venus in the celestial sphere was taken very seriously by
the mesoamerican cultures~\citep[e.g.][]{galindo09}. Its synodic period
in days is:

\begin{equation}
  584 = 8 \times 73 = 2^3 \times 73.
\label{period-venus}
\end{equation}

\noindent In
principle this sounds as a relevant period to take into account because
the number \( 73 \) appears in this prime decomposition (as it happens for
the full solar year of \( 365 \) days expressed in~\eqref{prime365}).  
As such, the combination of equations~\eqref{period-venus} and~\eqref{prime365}
yields:

\begin{displaymath}
  \frac{ 365 }{ 5 } = \frac{ 584 }{ 8 } = 73,
\end{displaymath}

\noindent so that a Venus basic calendar relation (Kepler's third law)
is given by:

\begin{equation}
  365 \times 8 = 584 \times 5,
\label{venus-golden-rule}
\end{equation}

\noindent i.e., \( 8 \) solar years correspond exactly to \( 5 \) Venus
synodic periods.  Since the fundamental number \( 13 \) does not appear on the
prime decomposition of either \( 365 \) or \( 584 \) it is possible to
force it by multiplying equation~\eqref{venus-golden-rule} by \( 13 \),
i.e.:

\begin{gather*}
  13 \times 8 \times 365 = 13 \times 5 \times 584, \\
\intertext{so that:}
  \left( 2 \times 52 \right) \times 365 = \left( 2 \times 73 \right) \times
  260,
\end{gather*}

\noindent which yields the mesoamerican basic relation~\eqref{golden-rule}
after multiplication by the \( 1/2 \) monad.  In other words, the
necessity of introducing the number \( 13 \) into Venus basic calendar
relation expressed in equation~\eqref{venus-golden-rule} yields the
mesoamerican basic relation~\eqref{fuego-nuevo} and naturally introduces
the \( 260 \) day count.

\section{Uniqueness of the 260 day cycle}
\label{uniqueness}

\begin{figure*}
  \includegraphics[width=0.99\textwidth]{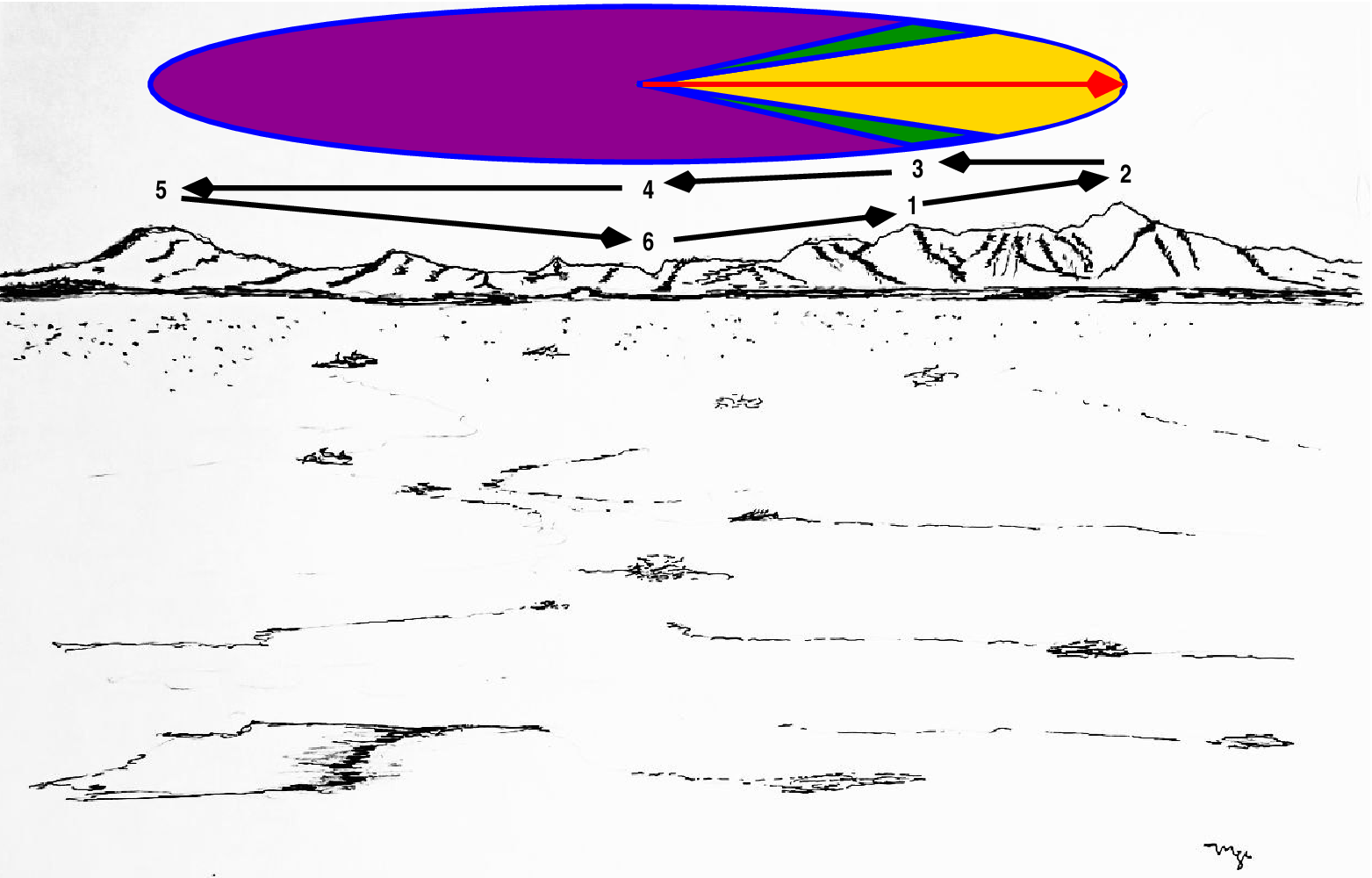}
\caption{The Figure shows a mountain range with a north-south orientation.
The sun rises or sets through the mountains between the two solstice
days represented by numbers 2 and 5 in the Figure.  At position 1, an
archaeoastronomical alignment occurs.  From it, after a certain number
of days, the solstice 2 is reached and after the same number of days,
the alignment repeats once more.  After this, the Sun continues to move
towards the equinox in point 4, reaching the solstice 5 and then coming
back to the other equinox in point 6 until it finally completes the
full solar year cycle of \( 365 \) days in 1.  The circular diagram
represents in magenta the short calendar count which is the number of
days elapsed between position 3 \( \rightarrow \) 4 \( \rightarrow \)
5 \( \rightarrow \) 6 \( \rightarrow \) 1.   The addition of the green
and yellow areas represent the number of days (including in the count
the given solstice day represented by the red arrow) to complete the \(
365 \) full solar year.  The addition of the magenta and green areas
represent the number of days between both alignments that do not pass
through the given solstice day and constitute an approximate solar count
\( \lesssim 365 \) days.  The yellow area are the extra days (including
the given solstice day represented by the red arrow) needed to complete
the full solar year count from the approximate count. The sketch was
kindly made by Marisela Mendoza.  }
\label{marisela}
\end{figure*}

 An archaeoastronomical alignment centered on a single given solstice day
consists on the alignment of a structure with the sunrise or sunset at
some particular dates.  To visualise this, Figure~\ref{marisela} shows a
mountain range with a north-south orientation in the horizon of a desert.
Over the full solar year, the sun rises/sets at some particular point
within the mountains  between the two solstices, represented by the
numbers 2 and 5 in the Figure.  If one requires an archaeoastronomical
alignment to occur at points 1 and 3, these two are symmetrical with
respect to the given solstice 2.  In other words, if an alignment occurs
at 1, then after a certain number of days the solstice 2 is reached
and then after the same number of days the alignment repeats again.
The sun continues to move over the year towards point 4 which represents
an equinox until it reaches the other solstice at point 5.  On its way
back, the sun reaches the other equinox at point 6, until the cycle is
completed at point 1.

  Archaeoastronomical alignments performed in ancient mesoamerica implied
the existence of a short calendar count, represented by the magenta area
in the circular passage of time (the number of days between positions 3 \(
\rightarrow \) 4 \( \rightarrow \) 5 \( \rightarrow \) 6 \( \rightarrow
\) 1).  The remaining number of days, represented by the addition of the
green and yellow areas in the circular diagram plus the given solstice day
depicted by the red arrow complete the full solar
cycle of \( 365 \) days.

  In order to account mathematically for the short calendar count, an
approximate calendar count \( \lesssim 365 \) days is required and is
represented in the diagram as the sum of the magenta and green areas.
The extra days to complete the full solar year count of \( 365 \) days
are represented by the yellow area on the diagram plus the given solstice
day depicted by the red arrow.

  Let us now find out whether the construction of an approximate calendar
with a corresponding short calendar is a unique choice.  To do so, let us
take as a reference point the sunrise or sunset of any of the June or 
December solstices and make the following definition for a short calendar:

\begin{definition}[Short calendar count]
\label{def01}

Let \( N = 365 \) represent the total number of days in a full solar year.
If \( n \lessapprox N \) is a natural number of an approximate solar
year in days and if the integer number \( d \) is a divisor of \( n \),
then an approximate solar count \( n \) satisfies a basic mesoamerican
relation (Kepler's third law) with the full solar count of \( N=365 \)
days if the integer quantity:

\begin{equation}
   q := n \left( \frac{ d - 2 }{d } \right) = n - \frac{ 2 n }{ d },
\label{testN}
\end{equation}

\noindent is divisible by any of the divisors of \( 365 \), i.e. if the
expression in equation~\eqref{testN} is divisible by \( 5 \) and/or \( 73\). 
To restrict
more the possible \(q\)'s and in order to be in harmony with a base \( 20 \)
numerical system used by the ancient mesoamericans, let us also impose the
quantity \( q \) to be divisible by \( 20 \) so that the number of \( q \)
days can be organised in sets of twenty (veintenas).  The accumulation of 

\begin{equation}
  q_{20} := q / 20,
\label{q_20}
\end{equation}

\noindent veintenas corresponds to the full number of days \( q \) in the
short calendar.  This  \emph{fundamental mesoamerican number} is
the key for cycling up veintenas that reach the number \( q \).  Note that
since we required \( q \) to be divisible by \( 20 \), it is also divisible
by \( 4 \) and \( 5 \).

\end{definition}

\begin{remark}
  Since \( q \) is divisible by \( 5 \), then let the integer number:

\begin{equation}
  q_5 := q / 5,
\label{q_5}
\end{equation}

\noindent so that the corresponding \emph{extended mesoamerican relation} 
is given by \( 5 = q / q_5 = 365 / 73 \).  In other words:

\begin{equation}
  q \times 73 = q_5 \times 365.
\label{kepler5}
\end{equation}

\noindent Similarly, if
\( q \) is divisible by \( 73 \), then define the integer number 

\begin{equation}
 q_{73} := q / 73,
\label{q_73}
\end{equation}

\noindent so that the corresponding \emph{extended mesoamerican
relation} is \( 73 = q / q_{73} = 365 /5 \), i.e.:

\begin{equation}
q \times 5  =  q_{73} \times 365.
\label{kepler73}
\end{equation}

  For this particular case, the restriction of \( q \) being also
divisible by \( 20 \) makes the approximate solar count \( n \) not to
be quite close to \( 365 \) and so, in what follows we will not consider
any divisibility of \( q \) by \( 73 \).
\end{remark}

\begin{remark}

  If \( n \) is a multiple of \( 5 \), relation~\eqref{testN} is naturally
divisible by \( 5 \).  This divisibility also occurs if \( 7 \) is the
last digit of a particular divisor \( d \) of the approximate calendar \(
n \), i.e. if \( d = 7 \times 10^0 + a_1 \times 10^1 + a_2 \times 10^2
+ \ldots \), for some natural numbers \( a_1 \), \( a_2 \), \ldots, \,
since \( d - 2  = 5 \times 10^0 + a_1 \times 10^1 + a_2 \times 10^2 +
\ldots\), which is indeed divisible by \( 5 \). Analogously, if \( 2 \)
is the last digit of a particular divisor \( d \), the statement holds.
Note that to fulfil the requirements of Definition~\ref{def01} the right-hand
side of equation~\eqref{testN} should also be divisible by \( 20 \).

\end{remark}

  As an example, let us choose \( n = 364 = 2^2 \times 7 \times
13 \). It then follows that the choice \( d = 7 \) satisfies the
requirements of Definition~\ref{def01}, with \( q = 260 \) and a basic mesoamerican relation given
by equation~\eqref{fuego-nuevo}.  Since \( q = 260  \)  is divisible
by \( 20 \) then the fundamental mesoamerican number is \( q_{20}
= 13\).  This result shows the relevance of number \( 13 \) for ancient
mesoamerican cultures and how it was constructed.

 In what follows, it will be useful to define the following natural
numbers:

\begin{gather}
  q_c:= n - q = \frac{ 2 n }{ d } = \frac{ 2 q }{ d - 2 },
  						\label{q_c}  \\
  q_{d-2} := \frac{ q_c }{ 2 } = \frac{ n - q }{ 2 } = \frac{ n }{ d }
           = \frac{ q }{ d - 2 },
	   					\label{q_b-2} \\
  \begin{split}	   					
  ( 2 q_5 )_c :=& n - 2 q_5 = n - \frac{ 2 }{ 5 } q,
  					\\
              =& \frac{ n }{ 5 } \left(
    \frac{ 3 d + 4 }{ d } \right) = \frac{ q }{ 5 } \left( \frac{ 3 d + 4
    }{ d - 2 } \right),
  \end{split}
    						\label{(2q_5)_c}\\
  n_c := n_\text{nem} := N - n = 365 - n,
  						\label{n_c}	\\
  \begin{split}
  n^* :=& \frac{ n_c - 1 }{ 2 } = \frac{ n_\text{nem} - 1 }{ 2 } 
      = \frac{ N - n - 1 }{ 2 } 
      					\\
      =& \frac{ 365 - n - 1 }{ 2 } = 182 - \frac{
      n }{ 2 }.
  \end{split}
      						\label{n*}
\end{gather}

\noindent where \( q_c \) is the number of days to reach \( n \) from \(
q \), \( q_{d-2} \) represents half \( q_c \), \( (2 q_5)_c \) stands for
the number of days to reach the number \( n \) from the value \( 2 q_5
\), \( n_c \) measures the deviation of days from the full solar year \(
N \), i.e. it is the number of days required to reach \( N \) from \( n \)
and \( 2 n^* \) is  the number \( n_c \) minus \( 1 \) (the given solstice day).

\begin{figure}
  \psfrag{A}{\textcolor{blue}{\huge $\mathbf{q}$}}
  \psfrag{B}{\textcolor{blue}{\Large $\mathbf{n\!^*}$}}
  \psfrag{D}{\textcolor{blue}{{\huge $\mathbf{q}$}{\huge $_{
    {}_{\mathbf{d-2}} }$}} }
  \includegraphics[width=0.47\textwidth]{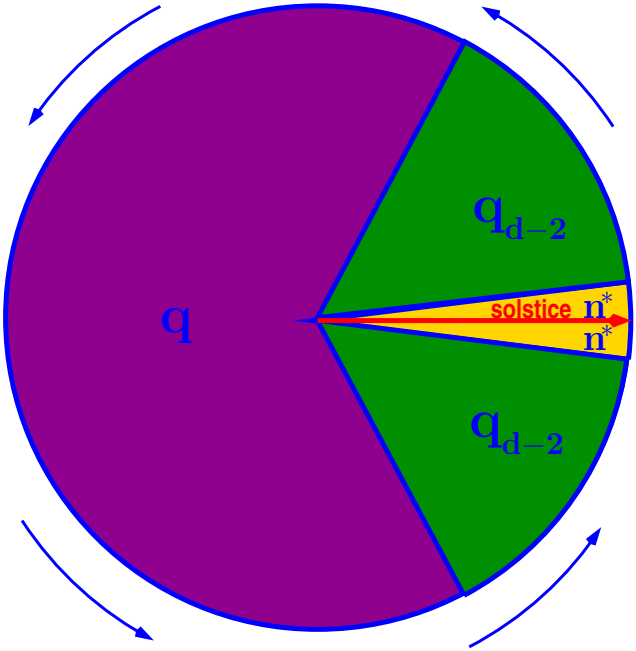}
\caption{The figure shows a circular schematic representation of Type-A
  archaeoastronomical alignments occurring on a full
  \( 365 \) day solar calendar.  The alignments occur \( q_{d-2} \)
  days before and after a given solstice.  The remaining \( q \) number
  of days between both alignments represent a short calendar count.
  The full solar year of \( 365 \) days is the sum: \( q + 2 q_{d-2} +
  2 n^* \) plus \( 1 \) (the chosen solstice day).
}
\label{type-A-alignment}
\end{figure}

  From the previous definitions we can construct two types of
archaeoastronomical alignments:

\begin{definition}[Type-A alignments]
\label{type-a-definition}
   Using as a base the concept of a short calendar count \( q \) in
Definition~\ref{def01} and the fact that:

\begin{equation}
  \begin{split}
  N = n + n_c =& q + q_c + n_c,
  		\\
	      =& q + 2 q_{d-2} + 2 n^* + 1,
  \end{split}
\label{type-A-motivation}
\end{equation}

\noindent it is natural to define two archaeoastronomical alignments
on a full solar year that occur \( q_{d-2} + n^* \) days before and
after a given solstice day.  The remaining number of days between each
alignment is \( q \).  This type of alignment is shown pictorically in
Figure~\ref{type-A-alignment}.

\end{definition}

\begin{figure}
  \psfrag{A}{\textcolor{blue}{\huge $\mathbf{(2q_{{}_5})_{{}_c}}$}}
  \psfrag{B}{\textcolor{blue}{\Large $\mathbf{n\!^*}$}}
  \psfrag{D}{\textcolor{blue}{{\huge $\mathbf{q_{{}_5}}$}} }
  \includegraphics[width=0.47\textwidth]{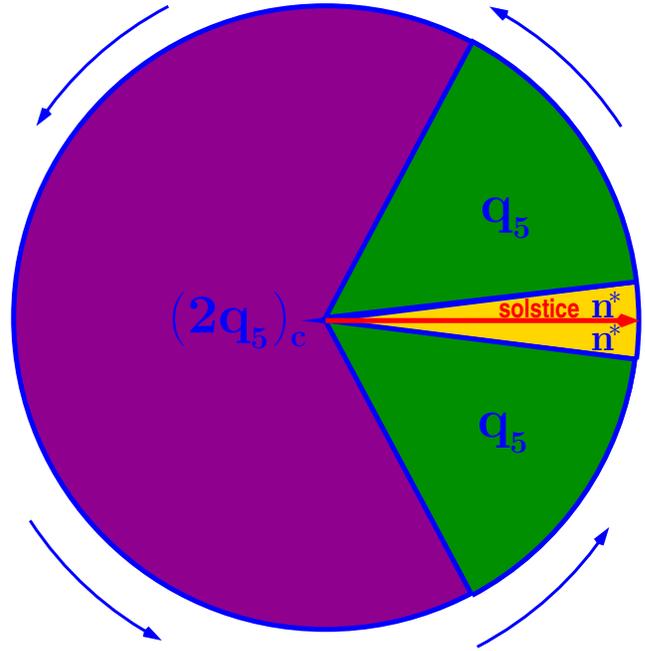}
\caption{The figure shows a circular schematic representation of Type-B
  archaeoastronomical alignments that occur on a full \( 365 \) day solar
  calendar. The alignments take place \( q_5 = q / 5\) days before and
  after a given solstice, with \( q \) the short calendar count. The
  remaining number of days \( (2q_5)_c \) can be thought of  as a 
  short calendar count for Type-B alignment types.  The full solar
  year of \( 365 \) days is the sum: \( (2q_5)_c + 2 q_{5} + 2 n^* \) plus
  \( 1 \) (the chosen solstice day).
}
\label{type-B-alignment}
\end{figure}

\begin{definition}[Type-B alignments]
\label{type-b-definition}
  Since the quantities \( q_{d-2} \) and \( q_5 \) appear
in a very symmetrical form  on the basic mesoamerican
relation~\eqref{kepler5}, instead of using \( q_{d-2} \) in
equation~\eqref{type-A-motivation}, let us use \( q_5 \) so that:

\begin{equation}
  \begin{split}
  N = n + n_c =& (2 q_5)_c + 2 q_5 + n_c,
  			\\
     	      =& (2 q_5)_c + 2 q_5 + 2 n^* + 1.
  \end{split}
\label{type-B-motivation}
\end{equation}

  This leads to the construction of two archaeoastronomical alignments
that occur \( q_5 + n^* \) days before and after a given solstice day.
The number of days to complete the full solar year \( N \) is \( ( 2 q_5
)_c \) plus 1 (the given solstice day). In this sense, the quantity \(
( 2 q_5 )_c \) can be taken as a short calendar count, analogous to
\( q \) for type-A alignments.  Type-B alignments are presented in
Figure~\ref{type-B-alignment}.

\end{definition}

  The previous two definitions imply directly the following result:

\begin{theorem}
  Type A and B archaeoastronomical alignments merge into a single one
when \( q_{d-2} = q_\text{5} \), which occurs for \( d = 7
\). In other words, a unique type of archaeoastronomical alignment can be
constructed with the requirements of Definition~\ref{def01} if the
approximate solar count \( n \) is divisible by \( d = 7 \). 
\end{theorem}

  Since the approximate solar count \( n = 364 \) with the choice 
\( d = 7 \) yields the short count \( q = 260 \), it follows that
only one type of archaeoastronomical alignment can be constructed with
these choices.

  I have written a Bourne Again Shell (bash) script code
named aztekas-calendar\footnote{The aztekas-calendar bash
script is \copyright~2023 Sergio Mendoza, licensed under
the GNU General Public License (GPL) and can be downloaded at
\url{https://aztekas.org/aztekas-calendar}.} in order to find out
natural numbers  \( \lessapprox 365 \) that satisfy the requirements of
Definition~\ref{def01}.  The results are shown in Table~\ref{table01}
for \( n \in [340,\ 364]  \).

\begin{table}
 \begin{tabular}{|r|r|r|r|r|r|r|r|l|}
\hline
\( n \) & \( q \) & \( q_{20} \) & \( d \) & \( q_{d-2} \) & \( q_5 \)  &
\( (2q_5)_c \) & Mesoamerican relation \\
\hline
340 &  300 &  15 &  17  &  20  &  60 &  220 &  300   \(\times\) 73 = 60 \(\times\) 365 \\
340 &  320 &  16 &  34  &  10  &  64 &  212 &  320   \(\times\) 73 = 64 \(\times\) 365 \\
344 &  340 &  17 &  172 &  2   &  68 &  208 &  340   \(\times\) 73 = 68 \(\times\) 365 \\
348 &  340 &  17 &  87  &  4   &  68 &  212 &  340   \(\times\) 73 = 68 \(\times\) 365 \\
350 &  280 &  14 &  10  &  35  &  56 &  238 &  280   \(\times\) 73 = 56 \(\times\) 365 \\
350 &  300 &  15 &  14  &  25  &  60 &  230 &  300   \(\times\) 73 = 60 \(\times\) 365 \\
350 &  340 &  17 &  70  &  5   &  68 &  214 &  340   \(\times\) 73 = 68 \(\times\) 365 \\
352 &  320 &  16 &  22  &  16  &  64 &  224 &  320   \(\times\) 73 = 64 \(\times\) 365 \\
360 &  120 &  6  &  3   &  120 &  24 &  312 &  120   \(\times\) 73 = 24 \(\times\) 365 \\
360 &  180 &  9  &  4   &  90  &  36 &  288 &  180   \(\times\) 73 = 36 \(\times\) 365 \\
360 &  240 &  12 &  6   &  60  &  48 &  264 &  240   \(\times\) 73 = 48 \(\times\) 365 \\
360 &  280 &  14 &  9   &  40  &  56 &  248 &  280   \(\times\) 73 = 56 \(\times\) 365 \\
360 &  300 &  15 &  12  &  30  &  60 &  240 &  300   \(\times\) 73 = 60 \(\times\) 365 \\
360 &  320 &  16 &  18  &  20  &  64 &  232 &  320   \(\times\) 73 = 64 \(\times\) 365 \\
360 &  340 &  17 &  36  &  10  &  68 &  224 &  340   \(\times\) 73 = 68 \(\times\) 365 \\
364 &  260 &  13 &  7   &  52  &  52 &  260 &  260   \(\times\) 73 = 52 \(\times\) 365 \\
364 &  360 &  18 &  182 &  2   &  72 &  220 &  360   \(\times\) 73 = 72 \(\times\) 365 \\
\hline
 \end{tabular}
\caption{The table shows different approximate solar calendar counts \( n
\) of a full solar year of \( N = 365 \) days, and their corresponding short counts \( q \) according to the
Definition~\ref{def01}. The meaning of the different columns is expressed
in equations~\eqref{q_c}-\eqref{(2q_5)_c}. The last column stands for the
corresponding basic Mesoamerican relation~\eqref{kepler5}.  Each line on
the table generates one Type-A and one Type-B archaeoastronomical alignment
as described in Definitions~\ref{type-a-definition} 
and~\ref{type-b-definition}.
}
\label{table01}
\end{table}

\begin{figure}
  \includegraphics[width=0.47\textwidth]{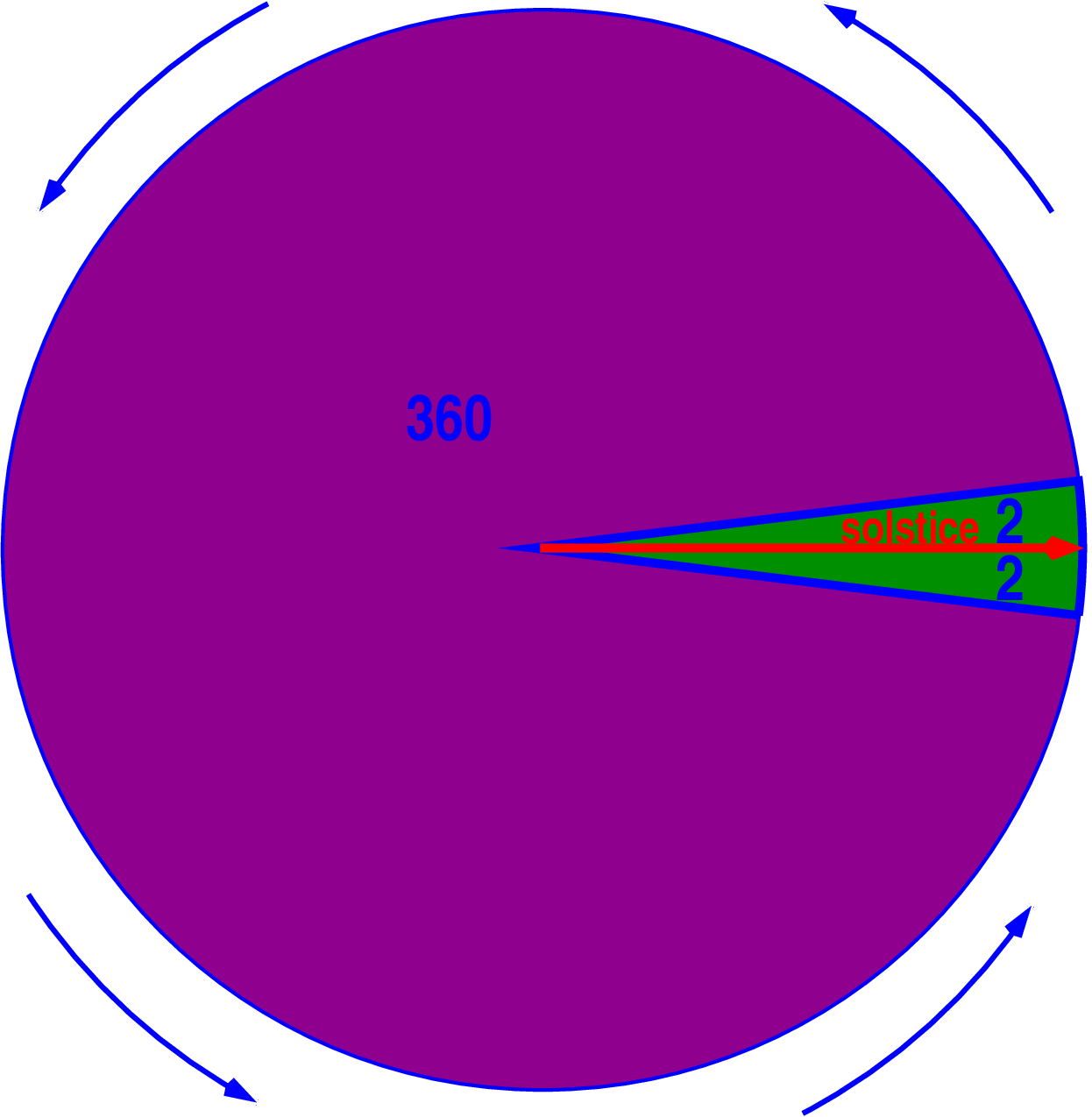}
  \caption{The Figure shows Type-A(\( n \! = \! 364,\, d \! = \! 182 \))
alignment which yields a \( q = 360 \) ``short'' calendar count.
Archaeoastronomical alignments occur 2 days before and after the given
solstice.  This construction can be used to align structures on the
solstice day and/or to the way a full solar year should be structured: \(
360 \) normal days plus \( 2 + 2 + 1 \) additional nemontemi days (counting
the solstice as one of the \( 5 \) special nemontemi days).
}
\label{360-short-type-A-figure}
\end{figure}

\begin{figure}
  \psfrag{219}{\textcolor{blue}{\huge $\mathbf{220}$}}
  \psfrag{73}{\textcolor{blue}{{\huge $\mathbf{72}$}} }
  \includegraphics[width=0.47\textwidth]{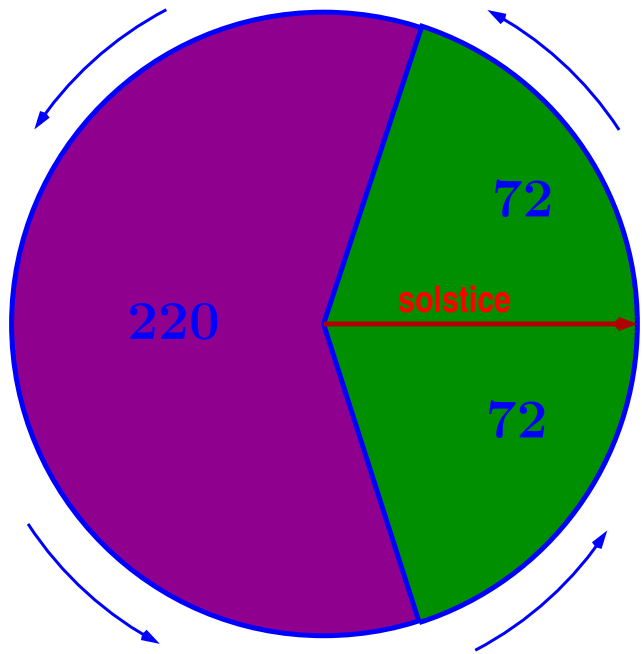}
  \caption{The Figure shows 
Type-B(\( n \! = \! 364, \, d \! = \! 182 \)) construction which has two
alignments occurring \( q_5 = 72 \) days before and after a given solstice.
These two sets of days plus their complement of \( 220 \) and the addition
of the given solstice day, complete the full solar year of \( 365 \) days.
}
\label{360-short-type-B-figure}
\end{figure}

  As expected one can construct a unique type-A and type-B short calendar
of \( q = 260 \) days based on an approximate solar calendar \( n = 364
\) days with the  fundamental mesoamerican number \( q_{d-2} = q_5 = 13
\) and with a basic mesoamerican relation~\eqref{fuego-nuevo}. 
For the same \( n = 364 \), the divisor \( d = 182 \) satisfies the
required test of Definition~\ref{def01} and so, a short count \( q=360 \) with
the fundamental basic mesoamerican number \( q_{20} = 18 \) 
can also be constructed with type-A and type-B alignments:

\begin{itemize}[leftmargin=*]
  \item Type-A(\( n \! = \! 364,\, d \! = \! 182 \)) \( \! \implies \! \) 
        (\(q \! = \!360\), \( q_{d-2} \! = \! q_{180} \! = \! 2, \, n^* \! =
	\! 0 \)).
	This is presented in Figure~\ref{360-short-type-A-figure}.
  \item Type-B(\( n \! = \! 364, \, d \! = \! 182 \)) \( \! \implies \! \) 
        (\( q_5 \! = \! 72,\, ( 2 q_5 )_c \! = \! 216,\,  n^* \! = \!
	0 \)).
	This is shown in Figure~\ref{360-short-type-B-figure}.
\end{itemize}

  It is important to note that with the archaeoastronomical requirements
mentioned in the definitions of equations~\eqref{testN} and~\eqref{q_20}, the
fundamental mesoamerican numbers \( 13 \) and \( 18 \) appear for the
short calendar counts of \( 260 \) and \( 360 \) days respectively,
both constructed with an approximate solar count of \( 364 \) days.

  Type-A(\( n \! = \! 364,\, d \! = \! 182 \)) alignment mentioned in the
previous list and shown in Figure~\ref{360-short-type-A-figure}  signals 
the way to construct a \( 360 \) normal day count plus \( 5 \) nemontemi
days.  In other words, the \( 360 \) day count
used by the ancient mesoamericans represents a short calendar count of a \(
364 \) approximate solar year.  The \( 365 \) day count is reached by adding
\( 5 \) nemontemi days.  This \( 360 + 5 \) day count cannot be
taken into account for an archaeoastronomical alignment
occurring \( 2 \) days before and after a given solstice, since both
alignment would be quite indistinguishable from each other. However, if one
chooses to have a single alignment occurring at any given solstice, then
this type of calendar count can be used.  Furthermore, there is
no necessity any more to have a solstice, since the chosen day for the alignment
can be also any other particular date of importance on any specific day of
the year, e.g. a single equinox day.

  Type-B(\( n \! = \! 364, \, d \! = \! 182 \)) alignment mentioned in
the previous list and shown in Figure~\ref{360-short-type-B-figure}
is such that \( 72 \) days before and after a given solstice an
archaeoastronomical alignment occurs.  The separation of \( 220 \) days
between both completes the approximate solar count  of \( n = 364 \)
days. The full solar year is completed when the solstice day is added
to the sum.  This is the correct way to express archaeoastronomical
alignments such as the one presented in Section~\ref{archaeoastronomy}
for the Templo Mayor in Tenochtitlan, and so the common way of describing
these type of alignment as presented in Figure~\ref{teotihuacan-figure}
is wrong.

  From the Definition~\ref{type-a-definition}
of Type-A alignment shown in
Figure~\ref{type-A-alignment} it follows that 
the approximate count \( n = q + 2 q_{d-2} \), i.e. the
approximate count \( n \) is splitted in a set of \( q \) days plus a set
of \( 2 q_{d-2} \) days, which can be written in terms of \( n \) as:

\begin{equation}
  n = \frac{ 2  }{ d } \, n  + \frac{ d - 2 }{ d } \, n.
\label{n-of-n-type-A}
\end{equation}

  The previous equation shows how the number of days on the approximate
count has been partitioned in the green (first term on the right-hand
side of equation~\eqref{n-of-n-type-A}) and the magenta (second
term on the right-hand side of relation~\eqref{n-of-n-type-A}) of
Figure~\ref{type-A-alignment}.  In analogous way, we can divide the
short count \( q \) into two parts:

\begin{equation}
  q = \frac{ 2  }{ d-2 } \, q   +  \frac{ d-4 }{ d-2 } \, q.
\label{q-of-q-type-A}
\end{equation}

  Relations~\eqref{n-of-n-type-A} and~\ref{q-of-q-type-A} motivate
a fundamental division between Type-A alignments to occur.  As such,
we can make the following definition:

\begin{definition} 
\label{general-monads-type-A}

  For Type-A alignments a family of two monad Type-A fraction sets are
defined as:

\begin{gather}
  m^{(\text{A})}_{n1} := \frac{ 2 }{ d }, \qquad \text{and}, \qquad 
    m^{(\text{A})}_{n2} := \frac{ d - 2   }{ d }.
				\label{monads-general-a-n} \\
  m^{(\text{A})}_{q1} := \frac{ 2 }{ d -2  }, \qquad \text{and}, \qquad 
    m^{(\text{A})}_{q2} := \frac{ d - 4   }{ d - 2 }.
				\label{monads-general-a-q}
\end{gather}

\end{definition}

  As an example, let us use \( n = 364 \) with \( d = 7 \), so that:

\begin{equation}
  m^{(\text{A})}_{n1} = 2 / 7, \quad 
  m^{(\text{A})}_{n2} = 5 / 7,  \quad 
  m^{(\text{A})}_{q1} = 2 / 5, \quad
  m^{(\text{A})}_{q2} = 3 /5, 
\label{monads-d-7}
\end{equation}

\noindent which correspond to the Teotihuacan and Aztec monad fractions
described in Section~\ref{archaeoastronomy}.

  In a similar way it is possible to define two monad Type-B fraction sets
using the following relations:

\begin{gather}
  n = \frac{ 2 \left( d-2 \right) }{ 5 d } \,  n + 
      \frac{ 3 d + 4 }{ 5 d } \, n,
      				\label{n-of-n-typeB} \\
  (2 q_5)_c = \frac{ 2 \left( d-2 \right) }{ 3 d + 4 }   
   \, (2 q_5)_c  + \frac{ d + 8 }{ 3d + 4 }  \, (2 q_5)_c,
   				\label{q-of-q-typeB}
\end{gather}

\noindent which are motivated by Definition~\ref{type-b-definition}
presented in Figure~\ref{type-B-alignment}.

\begin{definition} 
\label{general-monads-type-B}

  For Type-B alignments a family of two monad Type-B fraction sets are
defined as:

\begin{gather}
  m^{(\text{B})}_{n1} :=   \frac{ 2 \left( d-2 \right) }{ 5 d }, 
         \qquad \text{and}, \qquad 
    m^{(\text{B})}_{n2} := \frac{ 3 d + 4 }{ 5 d }.
				\label{monads-general-b-n} \\
  m^{(\text{B})}_{(2q_5)_c1} :=  \frac{ 2 \left( d-2 \right) }{ 3 d + 4 },
  \quad \text{and}, \quad 
    m^{(\text{B})}_{(2q_5)_c2} :=  \frac{ d + 8 }{ 3d + 4 }.
\end{gather}

\end{definition}

 From all explorations I performed for Type-A and Type-B monads,
all fractions produced seem not useful, except for the case \( d = 7 \) in
which both Type-A and Type-B monads converge to relations~\eqref{monads-d-7}.

  From the results of Table~\ref{table01} and using the fact that
Figure~\ref{360-short-type-A-figure} motivates the idea of a \( 360 + 5 \)
solar count, it is quite interesting to see that an approximate \( n = 360
\) days count can be chosen.  This choice produces the following
alignments:

\begin{itemize}[leftmargin=*]
  \item Type-A(\( n \! = \! 360,\, d \! = \! 3\)) \( \! \implies \! \) 
        (\(q \! = \!120\), \( q_{d-2} \! = \! q_{1} \! = \! 120, \, n^* \! =
	\! 2 \)).
  \item Type-B(\( n \! = \! 360, \, d \! = \! 3 \)) \( \! \implies \! \) 
        (\( q_5 \! = \! 24,\, ( 2 q_5 )_c \! = \! 312,\,  n^* \! = \!
	2 \)).
  \item Type-A(\( n \! = \! 360,\, d \! = \! 4\)) \( \! \implies \! \) 
        (\(q \! = \!180\), \( q_{d-2} \! = \! q_{2} \! = \! 90, \, n^* \! =
	\! 2 \)).
  \item Type-B(\( n \! = \! 360, \, d \! = \! 4 \)) \( \! \implies \! \) 
        (\( q_5 \! = \! 36,\, ( 2 q_5 )_c \! = \! 288,\,  n^* \! = \!
	2 \)).
  \item Type-A(\( n \! = \! 360,\, d \! = \! 6 \)) \( \! \implies \! \) 
        (\(q \! = \!240\), \( q_{d-2} \! = \! q_{4} \! = \! 60, \, n^* \! =
	\! 2 \)).
  \item Type-B(\( n \! = \! 360, \, d \! = \! 6 \)) \( \! \implies \! \) 
        (\( q_5 \! = \! 48,\, ( 2 q_5 )_c \! = \! 264,\,  n^* \! = \!
	2 \)).
  \item Type-A(\( n \! = \! 360,\, d \! = \! 9 \)) \( \! \implies \! \) 
        (\(q \! = \!280\), \( q_{d-2} \! = \! q_{7} \! = \! 40, \, n^* \! =
	\! 2 \)).
  \item Type-B(\( n \! = \! 360, \, d \! = \! 9 \)) \( \! \implies \! \) 
        (\( q_5 \! = \! 56,\, ( 2 q_5 )_c \! = \! 248,\,  n^* \! = \!
	2 \)).
  \item Type-A(\( n \! = \! 360,\, d \! = \! 12 \)) \( \! \implies \! \) 
        (\(q \! = \!300\), \( q_{d-2} \! = \! q_{10} \! = \! 30, \, n^* \! =
	\! 2 \)).
  \item Type-B(\( n \! = \! 360, \, d \! = \! 12  \)) \( \! \implies \! \) 
        (\( q_5 \! = \! 60,\, ( 2 q_5 )_c \! = \! 240,\,  n^* \! = \!
	2 \)).
  \item Type-A(\( n \! = \! 360,\, d \! = \! 18 \)) \( \! \implies \! \) 
        (\(q \! = \!320\), \( q_{d-2} \! = \! q_{16} \! =  \! 20, \, n^* \! =
	\! 2 \)).
  \item Type-B(\( n \! = \! 360, \, d \! = \! 18  \)) \( \! \implies \! \) 
        (\( q_5 \! = \! 64,\, ( 2 q_5 )_c \! = \! 232,\,  n^* \! = \!
	2 \)).
  \item Type-A(\( n \! = \! 360,\, d \! = \! 36 \)) \( \! \implies \! \) 
        (\(q \! = \!340\), \( q_{d-2} \! = \! q_{34} \! = \! 10, \, n^* \! =
	\! 2 \)).
  \item Type-B(\( n \! = \! 360, \, d \! = \! 36  \)) \( \! \implies \! \) 
        (\( q_5 \! = \! 68,\, ( 2 q_5 )_c \! = \! 224,\,  n^* \! = \!
	2 \)).
\end{itemize}

\begin{figure}
  \includegraphics[width=0.47\textwidth]{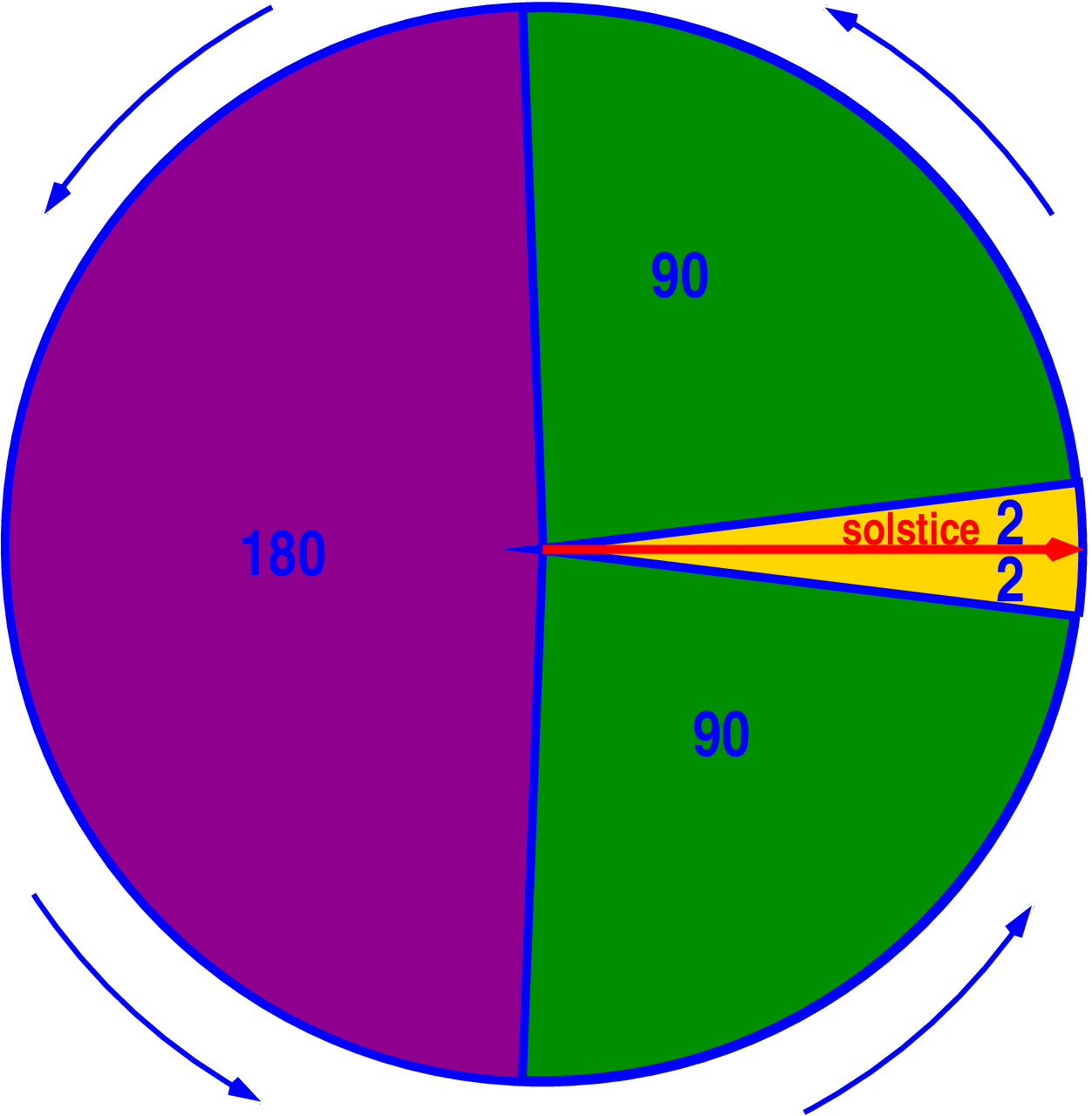}
  \caption{The Figure shows a circular calendar count of the full solar
year of \( 365 \) days. It corresponds to 
Type-A(\( n \! = \! 360,\, d \! = \! 4\)), so both alignments occur \( 92
\) days before and after a given solstice.  Thus, they both approximately
occur on the March and September equinoxes.  There are \( 180 \) remaining
days that added with \( 1 \) solstice day add up to reach 
the \( 365 \) days on a full solar year. 
}
\label{typeA_360-approx_180-short}
\end{figure}

  From the previous list, it is important to look at
the case Type-A(\( n \! = \! 360,\, d \! = \! 4\)) shown in
Figure~\ref{typeA_360-approx_180-short} since the archaeoastronomical
alignments occur \( 92 \) days before and after the given solstice.
In other words, both alignments occur quite close to the March and
September equinoxes.  This is the reason as to why some mesoamerican
monuments, e.g.  the pyramid of Kukulkan in Chichen Itza~\citep[see
e.g.][]{galindo09,galindo10}, present alignments on the equinoxes.

 The remaining Type-A and Type-B alignments presented in the previous list
may have some representations in mesoamerican cultures and should be
searched for in future archaeoastronomical/archaeological investigations.

\section{Discussion}
\label{discussion}

  This article began with the exploration of prime number decompositions
of the relevant counts of \( 260 \) and \( 365 \) days used in ancient
mesoamerica.  With this it was shown that if number \( 13 \) was forcibly
introduced into some important relations then a short count of \( 260 \)
days is needed.  On this exploration, it was also shown that the basic
mesoamerican calendar relation presented in equation~\eqref{fuego-nuevo}
is  a representation of Kepler's third law in synodic coordinates related
to the orbital period of the Earth about the Sun and an imaginary orbital
synodic period of \( 260 \) days.  There is no evidence, at least written,
that a knowledge of Kepler's laws of planetary motion were known to the
mesoamerican cultures. Nevertheless, it is surprising that such basic
calendar relation, constructed using basic arithmetical computations
related to the number of days on a full solar year and the short \(
260 \) count, is an indirect representation of Kepler's third law of
planetary motion of two imaginary planets: one with a perfect \( 260 \)
day circular periodic orbit and a perfect circular \( 365 \) day orbit of
the Earth about the Sun.  The key to build a mathematical concept of a short
calendar count was to merge all these concepts into two well known types of
archaeoastronomical alignments that occur at sunrise/sunset with respect to
a given solstice day, and were exemplified  in Section~\ref{archaeoastronomy}
with alignments occurring at Teotihuacan and Tenochtitlan.

  With the explorations mentioned in the previous paragraph it was then
possible to give a precise mathematical definition of a short calendar
in Section~\ref{uniqueness}. The definition uses the concept of an
\emph{approximate calendar count} and takes into account the existence of
a generalised basic calendar mesoamerican relation (Kepler's third law).
To be in harmony with a base 20 numerical system, the short calendar count
was postulated to be divisible by \( 20 \).  The ratio between the short
calendar count and the number \( 20 \) makes it possible to group days
into veintenas (sets of twenty) so that after twenty times that ratio,
the short calendar count is completed. 

  The most relevant approximate solar calendar count found in this article
that satisfies the requirements of Definition~\ref{def01} was \( 364 \)
days and was presented pictorically in Figure~\ref{teotihuacan-figure},
with a short calendar count of \( 260 \) days.  The ratio of \( 260 \)
over \( 20 \) produces the very well known fundamental number \( 13
\) used by ancient mesoamericans.  The short calendar count occurs \(
52 \) days after a given solstice, exactly at one archaeoastronomical
alignment and ends \( 260 \) days after that signalled by another
archaeoastronomical alignment.  When \( 52 \) days after this second
alignment have elapsed, the chosen solstice day is reached again.
The full solar \( 365 \) day count is reached by adding the chosen
solstice day to the approximate count of \( 364 \) days.

 The general symmetry presented in an extended mesoamerican relation was
used to define two classes of alignments (or short calendar counts): 
Type-A and Type-B presented in 
Definitions~\ref{type-a-definition}-\ref{type-b-definition}.  
Then it was found
that with a short calendar of \( 360 \) days plus \( 5 \) nemontemi days
(one of these nemontemi days being the chosen solstice day), alignments 
occur \( 2 \) days
previous and subsequent to the given solstice day, or due to the approximate
accuracy of an alignment, one can take a single alignment occurring at
the solstice day or for that matter at any other reference day, e.g. an
single equinox.  For this type of short calendar, the ratio \( 360 \) over
\( 20 \) yields the important number \( 18 \) (which has the same relevance
as number \( 13 \) in the \( 260 \) short calendar count).  
Veintenas of \( 18 \) days were used in ancient mesoamerica 
for grouping days on the \( 360 \) count.

It was also shown that with the approximate solar count of \( 364 \)
days, another short calendar of \( 220 \) days can be constructed
since that number of days is required to complete the approximate solar
count of \( 364 \) days when two archaeoastronomical alignments occur
\( 72 \) days before and after a given solstice.  This represents a
completely unexpected way to account for these type of alignments
(exemplified in this article using the Templo Mayor at Tenochtitlan,
Mexico City)\footnote{As explained in Section~\ref{archaeoastronomy},
the usual explanation for these alignments is that they occur \( 73 \)
days before and after a given solstice, with the remaining number of
days being \( 219 \).  Although the reasoning is correct, the solstice
day needs to be inserted inside any of the periods of \( 73 \) days.
This erroneous way of partitioning the full solar year count of \( 365 \)
days was produced because of the innacuracy that occurs at astronomical
alignments: an archaeoastronomical alignment occurs always at a certain
date within a few days of approximation.}.

  It is important to note that there is no simple way to account in a
calendar for the full solar year which is \( 365 \) days plus \( 5 \)
hours + \(48 \) minutes + \( 46 \) seconds.  The Gregorian calendar we use
for our daily routinary life is not perfect, nor will it be any which we
try to construct.  As such, sooner or later a correction is to be
performed.  The mesoamerican counts of \( 365 \), \( 364 \), \( 360 \) and \(
260 \) days are no exception to this rule.  

  As mentioned in Section~\ref{numbers}, there is no written evidence
of corrections for leap years by ancient mesoamerican cultures.
Definition~\ref{def01} requires a perfect \( 365 \) day calendar count to
work in a calendar form so as to have two symmetrical archaeoastronomical
alignments to occur.  The only way to perform this on a calendar count is
to have it corrected in an analogous way as corrections are made on
leap years in the Gregorian calendar. In a sense, ancient mesoamerican
cultures were quite used to perform the corrections required every year
for the approximate calendar count of \( 360 \) days by the introduction of
the \( 5 \) nemontemi extra days\footnote{A possibility is that the leap
year correction of one day was added as an extra nemontemi day every
four years, but one has to take this statement with care since there is
no written evidence for this.}.

\section{Acknowledgements}
\label{acknowledgements}
This work was supported by a PAPIIT DGAPA-UNAM grant
IN110522. SM acknowledges support from CONACyT (26344).  The author
is grateful to Daniel Flores for fruitful discussions over the years
about the Teotihuacan culture.  The author dedicates this article to
Christos Emiliano Georgopolous Mendoza since all these thoughts started
while trying to show him (and failing to do so) that the prime number \(
13 \) is by no means a bad or unlucky number, but the contrary: it is
the fundamental mesoamerican number!


\bibliographystyle{aipauth4-2}
\bibliography{mendoza}

\end{document}